\documentclass[conference]{IEEEtran}
\usepackage{cite}
\usepackage{float}
\usepackage{wrapfig}

\ifCLASSINFOpdf
\usepackage[pdftex]{graphicx}
\else
\usepackage[dvips]{graphicx}
\fi

\usepackage[cmex10]{amsmath}
\usepackage{url}
\begin{document}
\title{Router deployment of Streetside Parking Sensor Networks in Urban Areas}

\author{\IEEEauthorblockN{Trista Lin, Herv\'e Rivano, Fr\'ed\'eric Le Mou\"el}
\IEEEauthorblockA{INRIA, Universit\'e de Lyon, INSA-Lyon, CITI-INRIA, F-69621, Villeurbanne, France\\
Email: trista.lin@inria.fr, herve.rivano@inria.fr, frederic.le-mouel@insa-lyon.fr}
}

\maketitle

%\begin{abstract}
%The deployment of urban infrastructure is very important for urban sensor applications. In this paper, we studied and introduced the deployment strategy of wireless on-street parking sensor networks. We defined a multiple-objective problem with four objectives, and solved them with real street parking map. The results show two sets of Pareto Front with the minimum energy consumption, sensing information delay and the amount of deployed routers and gateways. The result can be considered to provide urban service roadside unit or be taken into account while designing a deployment algorithm.   
%\end{abstract}

% \fi
\IEEEpeerreviewmaketitle

\subsection{Background}

The deployment of municipal wireless mesh networks is of vital importance to collect urban information. 
As the increasing parking search problem, we select the popular smart parking application and study its deployment problems. 
In particular, on-street parking applications require such an infrastructure for enabling large scale system at reasonable costs. 
Current existing smart parking projects, e.g., SFpark and SmartSantander, generally comprise in-ground sensors, routers and gateways.
The common in-ground parking sensor is magnetometer which only detects ferrous metals. The in-ground sensor, restricted by the soil medium and battery-limited energy, is reduced function device (RFD) which communicates directly with roadside overground routers/gateways. Router and gateway, both full function devices (FFD) and as parts of urban infrastructure, collect the parking available information from sensors to internet. Thus, the network connectivity and performance are strongly determined by the density of routers and gateways. In order to reach a maximum coverage, routers and gateways are generally installed at crossroads and manage sensors directly, then a mesh network is formed among routers and gateways. %\cite{Marks:deploymemt} highlights the most important objectives while deploying wireless sensor networks. 
\cite{Konstantinidis2010960, 6883589, Luo:jucs_17_14:optimization_of_gateway_deployment} all proposed a multi-objective evolutionary approach to aid the sensor deployment with a multi-objective problem. However, we do not see a real map to be considered in a multi-objective optimization. In this article, we are mainly interested in the map impact to the deployment of urban infrastructure for wireless parking sensor networks. We consider a multi-objective problem and optimize it using real on-street parking maps. Our contributions is to formulate the multi-hop mesh networks among urban infrastructure and solve them by a multi-objective optimization in order to aim at the smart street parking application. We also highlight the following insights on parking sensor network design: First, the total energy consumption of sensors is strongly related to the amount of intersections and of active FFDs. Second, the gateway deployment can be seen as a cluster problem. Once the selection of FFD is once, the cluster number will be the minimum amount of deployed gateway. Third, the sensing information delay is related to the average degree of the street parking graph. Thus, the complexity of city map is an important factor while building urban infrastructure.

\subsection{Methodology} 

To analyze the above problem, we propose a methodology to work on. We first define a multi-objective problem according to the network coverage and connectivity in a multi-hop mesh network. From these constraints, we can also get some parameters which are relevant to sensor's lifetime and sensing information delay. Next, we build two graphs, i.e, on-street parking network and wireless link set, to indicate the relationship between any two given intersections. Here, we consider two different maps which have an approximate parking area length but their street networks are completely different. Then, we take the adjacency matrices as the data inputs of our constraints and try to optimize them.

%\begin{enumerate}
%\item Define the multi-objective problem
%\item Take different parameters and solve this problem in sage math with cplex solver
%\item Build a graph $C=(N, E)$ from openstreetmap using Osmosis and Graphserver
%\item Get the pathloss matrix using distance model by pgRouting
%\item Trim manually the graph with the parking map on Lyon City's website\cite{Lyoncity}
%\item Calculate the results with real maps. 
%\item Different snapshots for different traffic cases
%\end{enumerate}

\subsection{Multi-objective problem}

Since a city map is built by intersections and roads, let $C = (N, E)$ be a city graph. For all $n_i$ and $n_j$ $\in N$, we give two binary variables to express the status of each crossroad: 

\begin{minipage}{\linewidth}
\begin{align*}
x_i &= \left\{ \begin{array}{rl} 1 & \mbox{if a FFD is installed in $n_i$} \\ 0 & \mbox{otherwise} \end{array} \right. \\
y_i &= \left\{ \begin{array}{rl} 1 & \mbox{if a gateway is installed in $n_i$} \\ 0 & \mbox{otherwise} \end{array} \right.
\end{align*}
\end{minipage}

The amounts of deployed FFDs and gateways are expressed in equations~\ref{eq:phi_x} and~\ref{eq:phi_y} respectively.

\begin{minipage}{\linewidth}
\begin{minipage}{.45\linewidth}
\begin{align}
\phi_x &= \sum_{i \in N} x_i \label{eq:phi_x}
\end{align}
\end{minipage}
\begin{minipage}{.45\linewidth}
\begin{align}
\phi_y &= \sum_{i \in N} y_i \label{eq:phi_y}
\end{align}
\end{minipage}
\end{minipage}

In such a case, there are ($\phi_x - \phi_y$) routers. For formulating these problems, we define some variables: $d_{i,j}$ is segment distance between $n_i$ and $n_j$ and $d_{max}$ is the maximum road length; $\rho_{i,j}$ is the sensor density uniformly on the road segment between $n_i$ and $n_j$; $f_i$ is packets aggregated from FFD in $n_i$ (packets/s). $\Gamma_{i,j}$ is the managed length on road segment (i,j) on the FFD in the intersection $n_i$. $k_{i,j}$ is the managed sensor amount of each road segment around the intersection $n_i$; $h_{i}$ is the path distance (hop count) from FFD in $n_i$ to the its corresponding gateway; $M_{ns}$ is the maximum sensor numbers per router. $M_{rt}$ is the maximum capacity of router (packets/s); $M_{gw}$ is the maximum capacity of gateway (packets/s); $M_{hop}$ is the maximum hop count. If there are parking places between $n_i$ and $n_j$, the managed lengths of both sides will be greater than the length of road segment, in equation~\ref{eq:sumofgamma}. And the managed length of each FFD can not be greater than the road segment, in equation~\ref{eq:gamma-dij}. The density of parking sensors varies according to the parking type. In equation~\ref{eq:maxsensor-ffd} the amount of managed sensors on each FFD can be calculated by the sum of each managed road length multiplied by sensor density. $M_{ns}$ is limited by the bandwidth allocation method. In equation~\ref{eq:xi-gammaij-dmax}, if $\Gamma_{i,j}$ is not zero, it implies that there must be a FFD in $n_i$ who manages one part of road segment $(i,j)$. $d_{max}$, calculated from real maps, is always greater than all road segments.

\begin{minipage}{\linewidth}
\begin{minipage}{.59\linewidth}
\begin{align}
\Gamma_{i,j} + \Gamma_{j,i}  \geq d_{i,j}                   \;\forall (i,j) \in E \label{eq:sumofgamma}\\
\Gamma_{i,j} \leq d_{i,j}                                   \;\forall (i,j) \in E \label{eq:gamma-dij}\\
\sum_{(i,j) \in E} \Gamma_{i,j} \; \rho_{i,j}  \leq M_{ns}  \;\forall i \in N \label{eq:maxsensor-ffd}
\end{align}
\end{minipage}
\begin{minipage}{.40\linewidth}
\begin{align}
x_i \geq \frac{\Gamma_{i,j}}{d_{max}}                       \;\forall j \in N \label{eq:xi-gammaij-dmax}
\end{align}
\end{minipage}
\end{minipage}

Once we get $\Gamma_{i,j}$, the amount of parking sensors managed by $n_i$ on the road segment $(i,j)$ is expressed in equation~\ref{eq:kij}. Then the power consumption is mainly related to the transmission power of each individual sensor. We assume that each sensor has a transmission power determined by the transmission distance so that the total energy consumption is shown in equation~\ref{eq:energy-consumption}. Obviously, to minimize $\Omega_{s.total}$, we shall optimize $k_{i,j}$, which is proportional to $\Gamma_{i,j}$. That is why more deployed FFDs can improve the energy efficiency of in-ground sensors. 

\begin{minipage}{\linewidth}
\begin{equation}
k_{i,j} = \left\lfloor \Gamma_{i,j} \rho_{i,j} \right\rfloor \;\; \forall (i,j) \in E\label{eq:kij} 
\end{equation}
\end{minipage}

\begin{minipage}{\linewidth}
\begin{equation}
\begin{split}
\phi_{\Omega} &= \sum \Omega_s = \sum_{n_i \in N} \sum_{s \in n_i} \Omega_{s.i}= \sum_{n_i \in N} \sum_{s \in n_i} O(P_{txmw.s.i}) \\
&= \sum_{n_i \in N} \sum_{s \in n_i} O(10^{\frac{P_{txdbm.s.i}}{10}}) = \sum_{n_i \in N} \sum_{s \in n_i} O(d_{s,i}^{\frac{1}{10}}) \\
&= \sum_{n_i \in N} \sum_{(i,j) \in E} \frac{1}{2} k_{i,j} (k_{i,j}+1) \\
\end{split}
\label{eq:energy-consumption}
\end{equation}
\end{minipage}

Then we give some binary variables to express the relationship between intersections in equations~\ref{eq:bij}--\ref{eq:gij} to establish a multi-hop network. 

\begin{minipage}{\linewidth}
\begin{align}
b_{i,j} &= \left\{ \begin{array}{rl} 1 & \mbox{if FFD in $n_j$ is parent of the one in $n_i$} \\ 0 & \mbox{otherwise} \end{array} \right.   \label{eq:bij}\\
a_{i,j} &= \left\{ \begin{array}{rl} 1 & \mbox{if FFD in $n_j$ is ancestor of the one in $n_i$} \\ 0 & \mbox{otherwise} \end{array} \right. \label{eq:aij}\\
g_{i,j} &= \left\{ \begin{array}{rl} 1 & \mbox{if gateway in $n_j$ manages router in $n_i$} \\ 0 & \mbox{otherwise} \end{array} \right.     \label{eq:gij}
\end{align}
\end{minipage}

Now, we start to define the multi-hop constraints. Once we decide the intersections to install FFD, we will choose some among them to install gateways and keep the remaining for routers, shown in equation~\ref{eq:gwFromFFD}. In equation~\ref{eq:parent-node-bii}, each node can not be its own parent. If $n_j$ is the parent of $n_i$, it is its ancestor as well in equation~\ref{eq:parent-node-bij}. However, it will not be the child of $n_i$ simultaneously in equation~\ref{eq:sum-bij}. In equation~\ref{eq:parent-node-bij-sum}, each router has only one parent node and each gateway has no parent.  

\begin{minipage}{\linewidth}
\begin{minipage}{.29\linewidth}
\begin{align}
y_i \leq x_i                          &\; \forall \;i \label{eq:gwFromFFD}    \\
b_{i,i} = 0                           &\; \forall \;i \label{eq:parent-node-bii}
\end{align}
\end{minipage}
\begin{minipage}{.69\linewidth}
\begin{align}
b_{i,j} \leq a_{i,j}                  &\; \forall \;(i,j) \in W \label{eq:parent-node-bij}\\
b_{i,j} + b_{j,i} \leq W_{i,j}        &\; \forall \;(i,j) \in W \label{eq:sum-bij}\\
\sum_{n_j \in N} b_{i,j} = x_i - y_i  &\; \forall \;(i,j) \in W \label{eq:parent-node-bij-sum}
\end{align}
\end{minipage}
\end{minipage}

In equation~\ref{eq:ancestor-node-aii}, each FFD is its own ancestor. In equations~\ref{eq:ancestor-node-aij-xi-xj}, if $a_{i,j}$ is equal to 1, it implies that there are FFDs installed both in $n_i$ and $n_j$. In equation~\ref{eq:ancestor-node-aij}, $n_i$ and $n_j$ can not be the ancestor of each other at the same time, i.e., the link is unidirectional.

\begin{minipage}{\linewidth}
\begin{minipage}{.29\linewidth}
\begin{align}
a_{i,i} = x_i                            &\;\; \forall \;i \label{eq:ancestor-node-aii}
\end{align}
\end{minipage}
\begin{minipage}{.69\linewidth}
\begin{align}
a_{i,j} \leq x_i, \; a_{i,j} \leq x_j     &\;\; \forall \;i, j \label{eq:ancestor-node-aij-xi-xj}\\
a_{i,j} + a_{j,i} \leq 1                 &\;\; \forall \;i \neq j \label{eq:ancestor-node-aij}
\end{align}
\end{minipage}
\end{minipage}

In equation~\ref{eq:gateway-node-gii}, each gateway is managed by itself. In equation~\ref{eq:gateway-node-gij-yj}, if $g_{i,j}$ is equal to 1, it implies that there is a gateway installed $n_j$. Since each gateway manages itself, it can not be the gateway of another gateway, shown in equation~\ref{eq:gateway-node-gij}. In equation~\ref{eq:gateway-node-gij-aij}, if the gateway in $n_j$ manages the router in $n_i$, the gateway is the ancestor of the router. In equation~\ref{eq:gateway-node-gij-sum}, each FFD is managed by exact one gateway.

\begin{minipage}{\linewidth}
\begin{minipage}{.32\linewidth}
\begin{align}
g_{i,i} = y_i                    &\; \forall \;i         \label{eq:gateway-node-gii}  \\
g_{i,j} \leq y_{j}               &\; \forall \;i         \label{eq:gateway-node-gij-yj} 
\end{align}
\end{minipage}
\begin{minipage}{.66\linewidth}
\begin{align}
g_{i,j} + g_{j,i} \leq 1         &\; \forall \;i \neq j  \label{eq:gateway-node-gij}  \\
g_{i,j} \leq a_{i,j}             &\; \forall \;i         \label{eq:gateway-node-gij-aij}  \\
\sum_{n_j \in N} g_{i,j} = x_i   &\; \forall \;i        \label{eq:gateway-node-gij-sum}
\end{align}
\end{minipage}
\end{minipage}

In equation~\ref{eq:multihop-bna}, if the router in $n_i$ is the child of the one in $n_j$ and the descendant of the one in $n_k$, the router in $n_j$ is also the descendant of the one in $n_k$. In equation~\ref{eq:multihop-ang}, if the router in $n_i$ is the descendant of the one in $n_j$ and also managed by the gateway $n_k$, the router in $n_j$ is also managed by the gateway in $n_k$.

\begin{minipage}{\linewidth}
\begin{align}
b_{i,j} + a_{i,k} \leq a_{j,k} + 1  &\; \forall \;i, j, k \label{eq:multihop-bna}\\
a_{i,j} + g_{i,k} \leq g_{j,k} + 1  &\; \forall \;i, j, k \label{eq:multihop-ang}
\end{align}
\end{minipage}

Hence, the hop limit can be counted by the amount of ancestors on each router in equation~\ref{eq:hop_count}. Also, $h_i$ is limited by the $M_{hop}$ in equation~\ref{eq:hop_max}. The average sensing information delay is calculated by the average hop count in equation~\ref{eq:average_delay}. Since $a_{i,j}$ is bounded by $b_{i,j}$, the problem $\phi_{h/x}$ is equivalent to $min(\sum\nolimits_{(i,j) \in E} b_{i,j})$ namely the amount of active wireless links.

\begin{minipage}{\linewidth}
\begin{minipage}{.45\linewidth}
\begin{align}
\sum_{n_j \in N} a_{i,j} = h_i  &\; \forall \;i \label{eq:hop_count}\\
h_i \leq M_{hop}  &\; \forall \;i \label{eq:hop_max}
\end{align}
\end{minipage}
\begin{minipage}{.54\linewidth}
\begin{align}
\phi_{h/x} = \frac{(\sum\limits_{n_i \in N} h_i)}{(\sum\limits_{n_i \in N} x_i)} \label{eq:average_delay}
\end{align}
\end{minipage}
\end{minipage}

Network capacity can be expressed in equation~\ref{eq:trafficload}, which is restricted by the maximum capacity of each FFD, i.e., $M_{gw}$ and $M_{rt}$. Packet generation rate $f_i$ depends on the vehicle's arrival and departure, and the proven packet interval is heavy-tailed which can be best described by Weibull distribution. 

\begin{minipage}{\linewidth}
\begin{align}
\sum_{n_i \in N} f_i \; a_{i,j} \leq M_{rt} + y_j \; (M_{gw} - M_{rt}) \;\; \forall \;j \label{eq:trafficload}
\end{align}
\end{minipage}

\begin{figure}[!t]
\begin{minipage}{.59\linewidth}
\centering
\includegraphics[width=\linewidth]{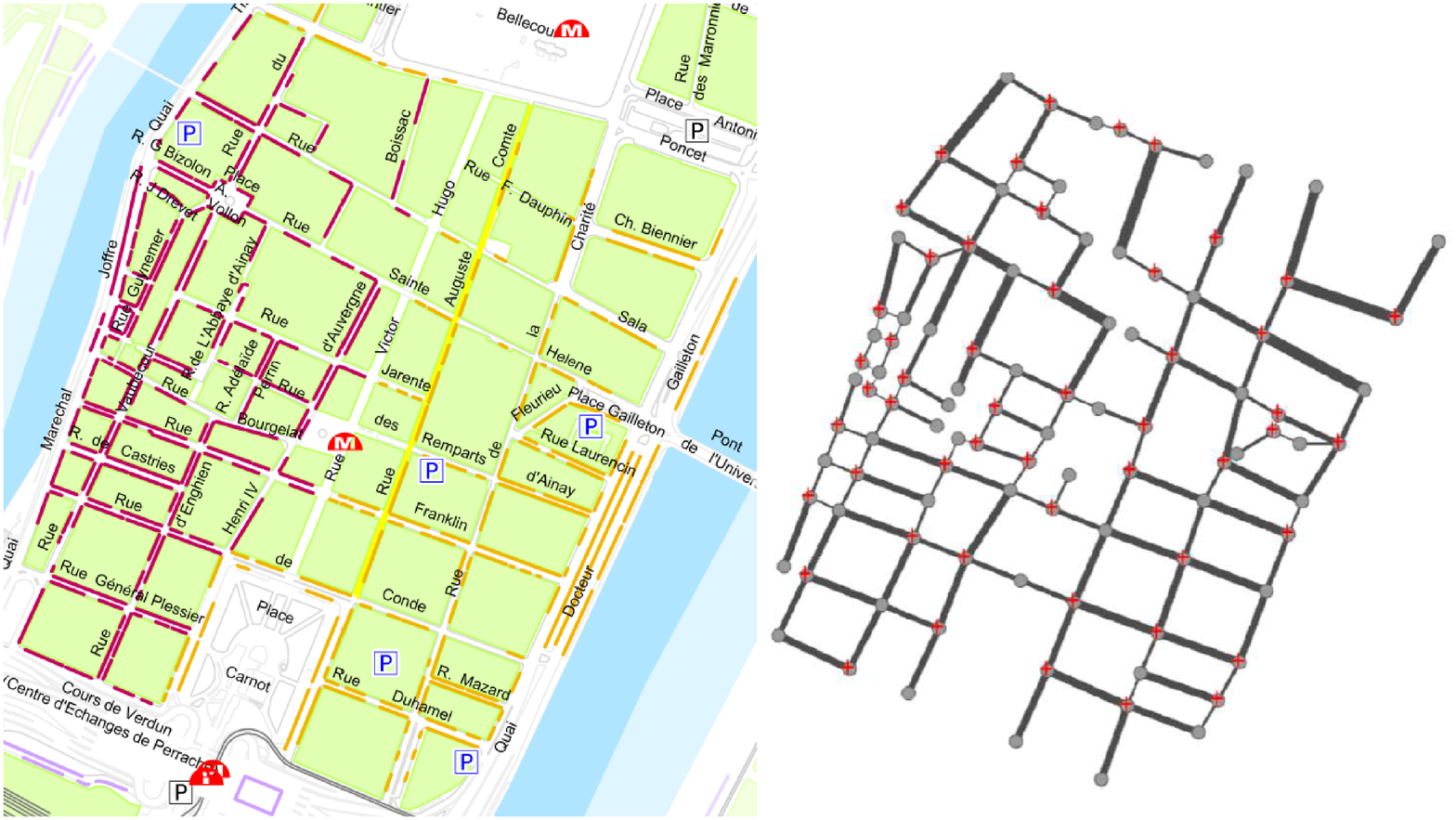}
\end{minipage}
\begin{minipage}{.40\linewidth}
\centering
\includegraphics[width=\linewidth]{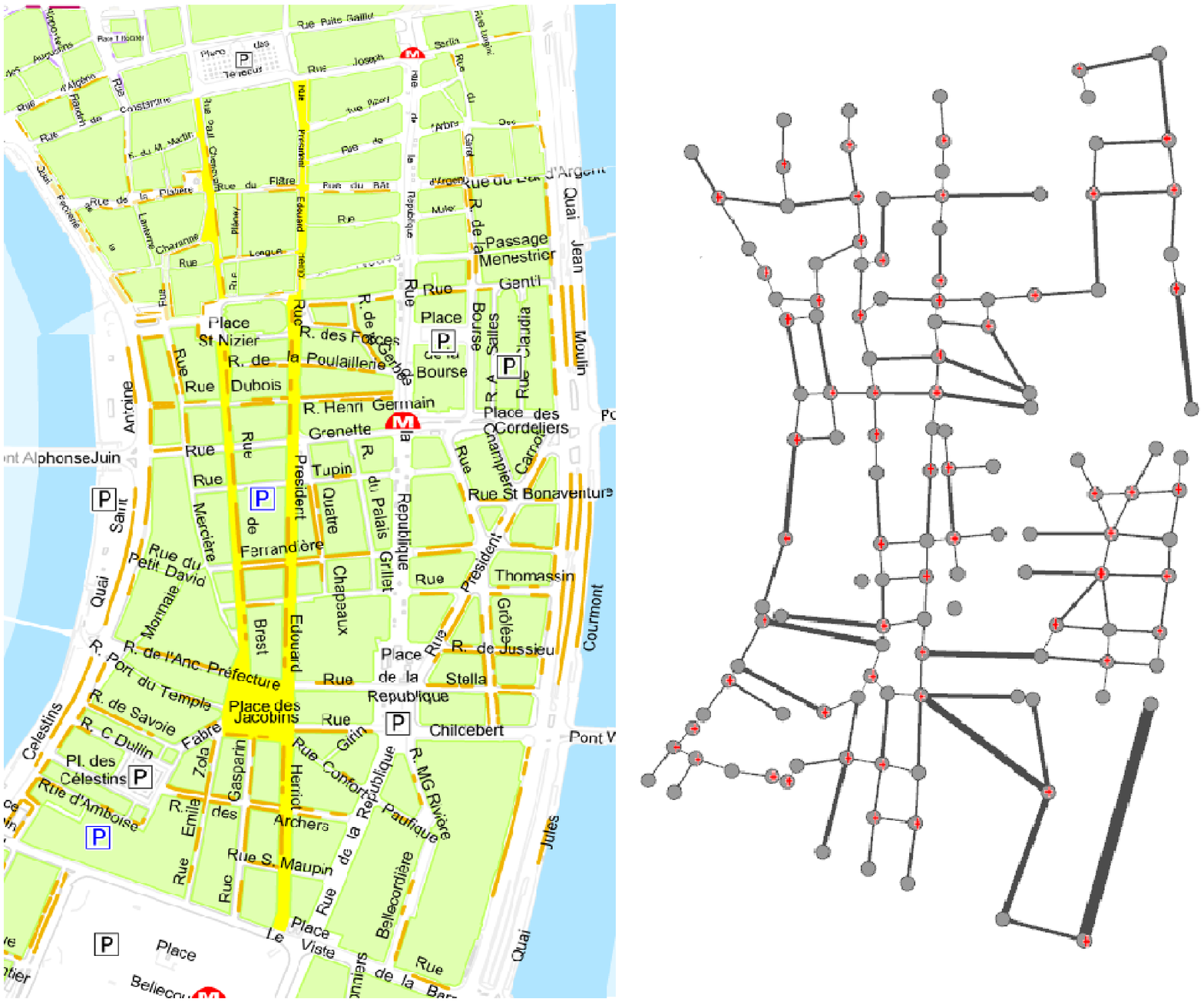} 
\end{minipage}
\caption{Map 1 (left)--Lyon on-street parking map between Place Bellecour and Place Carnot. The red dot indicates the selected intersections to install FFDs. Map 2 (right) --Lyon on-street parking map between Place Terreaux and Place Bellecour. The red dot indicates the selected intersections to install FFDs.}
\label{fig:ffd-deployment} 
%\end{figure}
%\begin{figure}[!t]
\begin{minipage}{.49\linewidth}
\centering
\includegraphics[width=\linewidth]{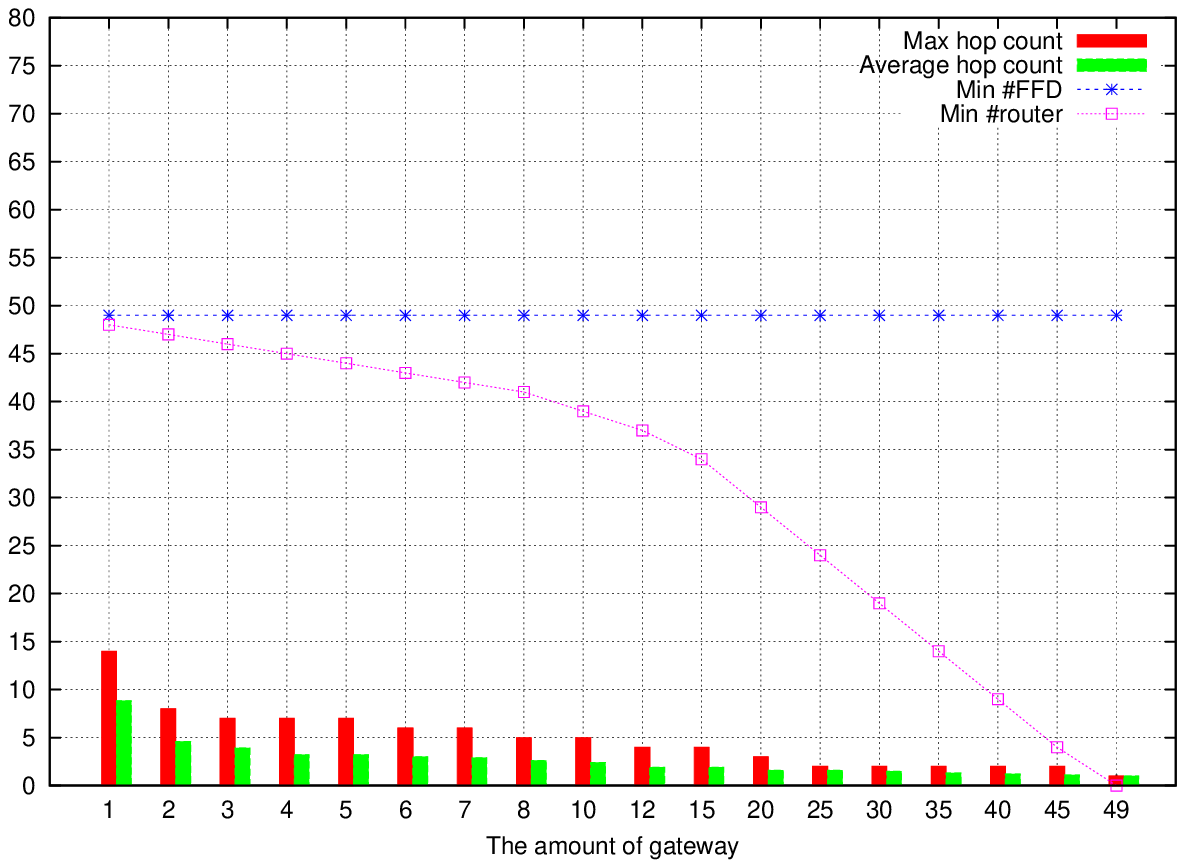}
\caption{Hop count v.s. the amount of deployed gateways in map 1}
\label{fig:gw-hop-map1}    
\end{minipage}
\hfill
\begin{minipage}{.49\linewidth}
\centering
\includegraphics[width=\linewidth]{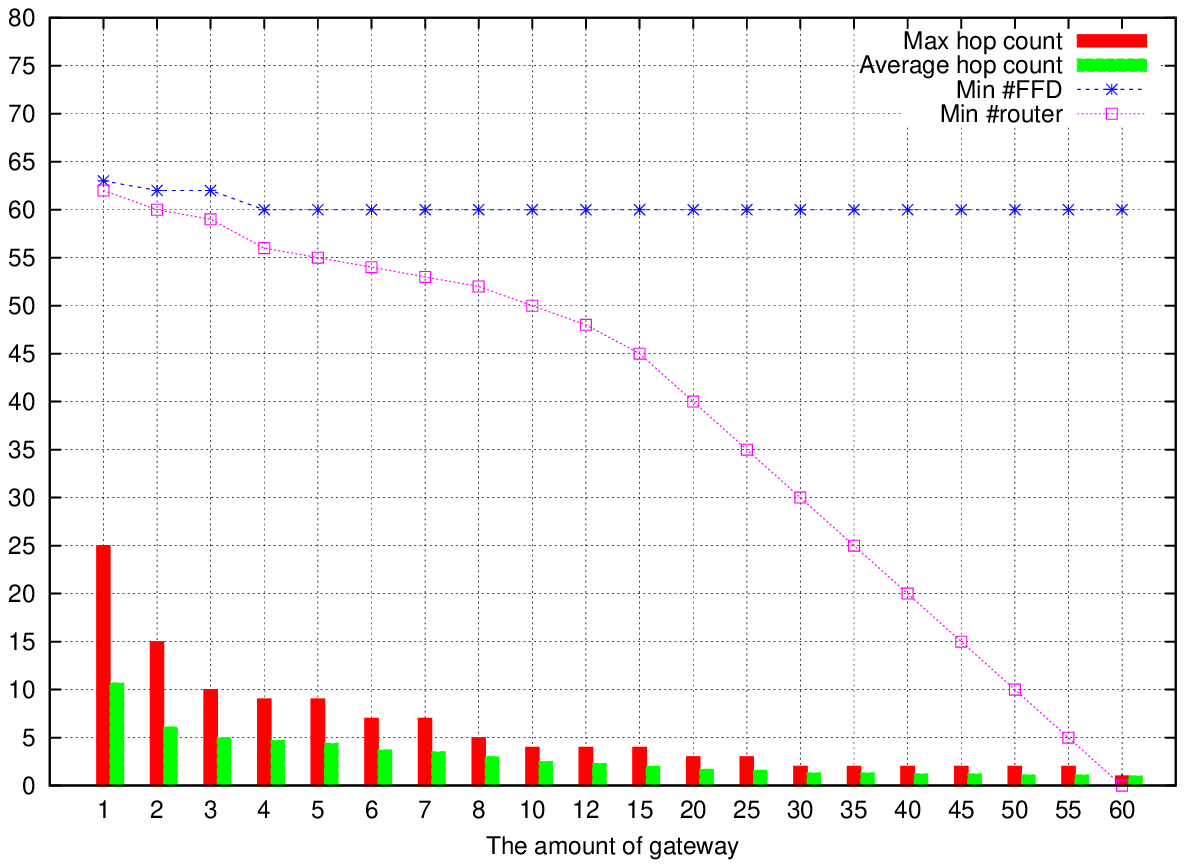}
\caption{Hop count v.s. the amount of deployed gateways in map 2}
\label{fig:gw-hop-map2}
\end{minipage}
\begin{minipage}{.49\linewidth}
\centering
\includegraphics[width=\linewidth]{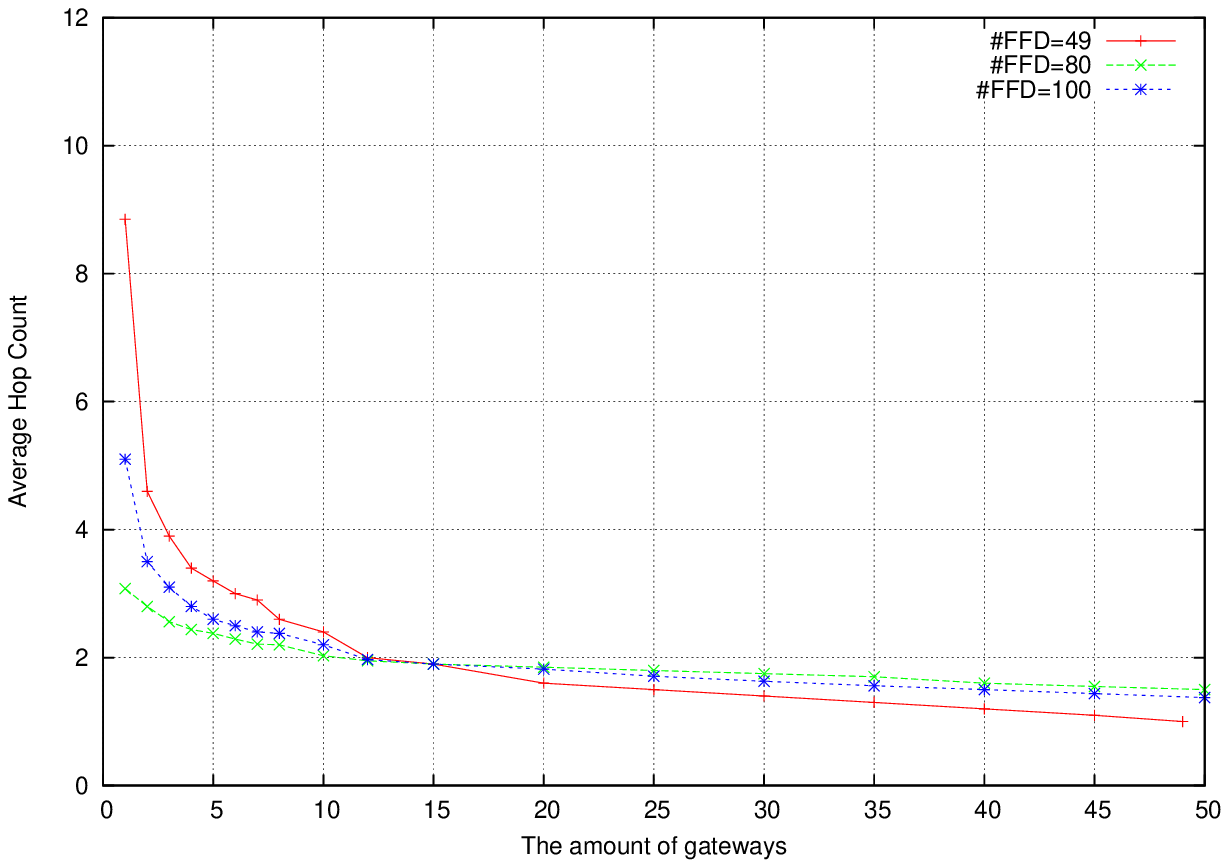}
\caption{Average hop count v.s. the amount of deployed gateways under the best, the mediocre and the worst cases for sensor lifetime in map 1}
\label{fig:gw-hop-ffd-map1} 
\end{minipage}
\hfill
\begin{minipage}{.49\linewidth}
\centering
\includegraphics[width=\linewidth]{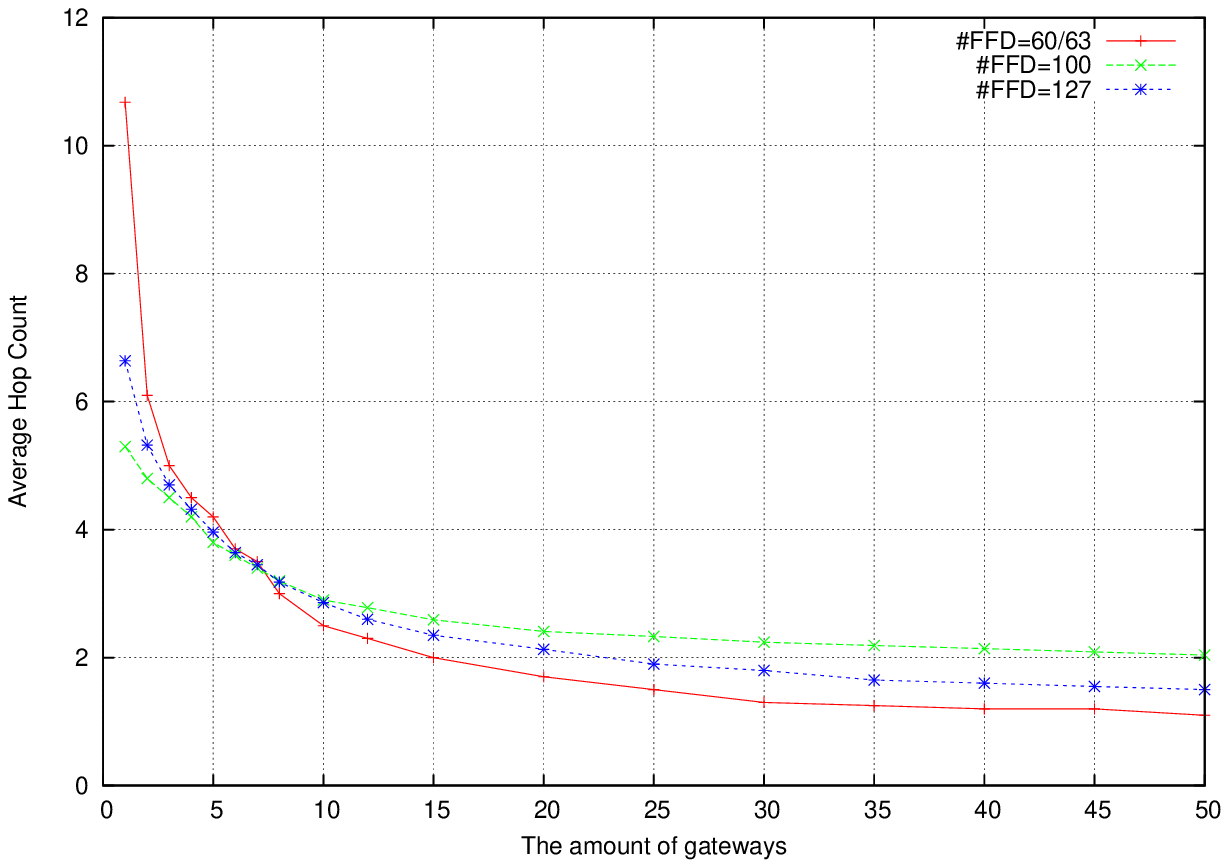}
\caption{Average hop count v.s. the amount of deployed gateways under the best, the mediocre and the worst cases for sensor lifetime in map 2}
\label{fig:gw-hop-ffd-map2}     
\end{minipage}
\end{figure}

With the above equations, we solve a multiple-objective problem $min(\phi_x, \phi_{\Omega}, \phi_y, \phi_{h/x})$ in Sage using the CPLEX solver. Two maps are retrieved in figure~\ref{fig:ffd-deployment}. Here we clipped two maps in order to have a very approximate total parking area length in both maps.

\subsection{Evaluation}

\paragraph{FFD deployment}
In figure~\ref{fig:ffd-deployment}, the red dots on the nodes are the optimally selected intersections to install FFDs. The minimum FFD amounts are 49 and 60 in map 1 and 2 respectively according to their map properties.

\begin{wrapfigure}{r}{0.45\linewidth}
  \begin{center}
    \includegraphics[width=\linewidth]{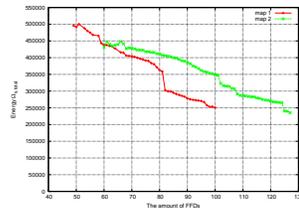}
  \end{center}
  \caption{Total sensor energy consumption v.s. the amount of deployed FFDs}
	\label{fig:gamma-x}
\end{wrapfigure}

\paragraph{Energy consumption}
From the set of $x_i$, we get $\Gamma_{i,j}$ and calculate the energy indicator by equation~\ref{eq:energy-consumption}. The first Pareto front result is shown in figure~\ref{fig:gamma-x}. The minimum FFD amounts are both more or less $50\%$. As the increasing of FFD amount, the energy depletion both reduced. Both of them take a dip when the deployed FFD amount accounts for $80\%$ of the intersections. Since the road segment length of map 1 is longer, more sensors have a longer transmission range which cause a higher transmission power. Thus, the maximum energy consumption in map 1 is higher than the one in map 2. While increasing the FFD amount, map 1, which has few intersections, can be covered faster than map 2 and the curve also drops more rapidly.

\paragraph{Gateway deployment}
The gateway deployment can be seen as a cluster problem. According to the amount of gateways, we divide all the FFDs into several partitions and then select one gateway from each of them. Figure~\ref{fig:gw-hop-map1} of map 1 shows that the amount of FFDs does not change no matter how many gateways are selected thanks to the grid-like topology. On the contrary, figure~\ref{fig:gw-hop-map2} of map 2 does not require additional FFD when there are more than 3 gateways. That is because those wireless links between the 60 FFDs form 4 clusters geographically. %When we select only one gateway in map 2, these 4 clusters require extra FFD in order to connect to the gateway. 
Thus, if we want to install gateway without creating additional wireless radio link in map 2, the minimum gateway amount will be 4. %Both figures~\ref{fig:gw-hop-map1} and~\ref{fig:gw-hop-map2} show that the hop count is reduced as the ratio of gateway to FFD increases.  

\paragraph{Sensing information delay}
The sensing information delay is generally related to the hop count as a result of the sleep-wake scheduling in wireless sensor networks. We take the hop count as the indicator of information delay. In figure~\ref{eq:energy-consumption}, we see the amount of FFDs really impacts to the energy consumption. Thus, we take three different amounts to stand for the worst, the mediocre and the best cases of sensor's lifetime, respectively. 
Figures~\ref{fig:gw-hop-ffd-map1} and~\ref{fig:gw-hop-ffd-map2} show the relationship between the hop count and the gateway amount. The worst energy case is shown in red and the best in green. The hop count decreases when the gateway amount increases. The mediocre case is selected at $80\%$ of the total intersections since there is a energy dip. Else. the hop count in map 1 converges slower than in map 2 because there are fewer intersections.

%\begin{figure*}[!t]
%\centering
%\subfloat[Case I]{\includegraphics[width=2.5in]{box}%
%\hfil
%\subfloat[Case II]{\includegraphics[width=2.5in]{box}%
%\label{fig_second_case}}
%\caption{Simulation results.}
%\label{fig_sim}
%\end{figure*}

%\begin{table}[!t]
%% increase table row spacing, adjust to taste
%\renewcommand{\arraystretch}{1.3}
% if using array.sty, it might be a good idea to tweak the value of
% \extrarowheight as needed to properly center the text within the cells
%\caption{An Example of a Table}
%\label{table_example}
%\centering
%% Some packages, such as MDW tools, offer better commands for making tables
%% than the plain LaTeX2e tabular which is used here.
%\begin{tabular}{|c||c|}
%\hline
%One & Two\\
%\hline
%Three & Four\\
%\hline
%\end{tabular}
%\end{table}

\subsection{Conclusion}

In this paper, we studied and introduced the wireless on-street parking sensor network from the viewpoint of system deployment. We highlighted some important factors and parametrized them in linear equations. To consider a more realistic urban environment, we retrieved two different parking maps with the same parking area length. The results shows two sets of Pareto Front which have different performances even both of them are merely from different blocks of the same city. The first Pareto Front shows the minimum energy consumption and the minimum amount of required FFDs. The second one shows the minimum hop count and the minimum amount of deployed gateways. The impact of real maps is shown by easily comparing any two figures. Since the urban sensor network attracts more and more attention to urban service, the gateway can also play the rule of road side unit. This way, the buffer size and the vehicle flow will also have to be considered in our future works.

\footnotesize{
\bibliography{IEEEexample}
\bibliographystyle{IEEEtran}
}

\end{document}